\documentstyle[twoside,psfig]{article}

\catcode`\@=11
\long\def\@makefntext#1{
\protect\noindent \hbox to 3.2pt {\hskip-.9pt  
$^{{\eightrm\@thefnmark}}$\hfil}#1\hfill}		

\def\thefootnote{\fnsymbol{footnote}}
\def\@makefnmark{\hbox to 0pt{$^{\@thefnmark}$\hss}}	
	
\def\ps@myheadings{\let\@mkboth\@gobbletwo
\def\@oddhead{\hbox{}
\rightmark\hfil\eightrm\thepage}   
\def\@oddfoot{}\def\@evenhead{\eightrm\thepage\hfil
\leftmark\hbox{}}\def\@evenfoot{}
\def\sectionmark##1{}\def\subsectionmark##1{}}

\oddsidemargin=\evensidemargin
\renewcommand{\thefootnote}{\fnsymbol{footnote}}

\newcounter{sectionc}\newcounter{subsectionc}\newcounter{subsubsectionc}
\renewcommand{\section}[1] {\vspace{12pt}\addtocounter{sectionc}{1} 
\setcounter{subsectionc}{0}\setcounter{subsubsectionc}{0}\noindent 
	{\tenbf\thesectionc. #1}\par\vspace{5pt}}
\renewcommand{\subsection}[1] {\vspace{12pt}\addtocounter{subsectionc}{1} 
	\setcounter{subsubsectionc}{0}\noindent 
	{\bf\thesectionc.\thesubsectionc. {\kern1pt \bfit #1}}\par\vspace{5pt}}
\renewcommand{\subsubsection}[1] {\vspace{12pt}\addtocounter{subsubsectionc}{1}
	\noindent{\tenrm\thesectionc.\thesubsectionc.\thesubsubsectionc.
	{\kern1pt \tenit #1}}\par\vspace{5pt}}
\newcommand{\nonumsection}[1] {\vspace{12pt}\noindent{\tenbf #1}
	\par\vspace{5pt}}

\newcounter{appendixc}
\newcounter{subappendixc}[appendixc]
\newcounter{subsubappendixc}[subappendixc]
\renewcommand{\thesubappendixc}{\Alph{appendixc}.\arabic{subappendixc}}
\renewcommand{\thesubsubappendixc}
	{\Alph{appendixc}.\arabic{subappendixc}.\arabic{subsubappendixc}}

\renewcommand{\appendix}[1] {\vspace{12pt}
        \refstepcounter{appendixc}
        \setcounter{figure}{0}
        \setcounter{table}{0}
        \setcounter{lemma}{0}
        \setcounter{theorem}{0}
        \setcounter{corollary}{0}
        \setcounter{definition}{0}
        \setcounter{equation}{0}
        \renewcommand{\thefigure}{\Alph{appendixc}.\arabic{figure}}
        \renewcommand{\thetable}{\Alph{appendixc}.\arabic{table}}
        \renewcommand{\theappendixc}{\Alph{appendixc}}
        \renewcommand{\thelemma}{\Alph{appendixc}.\arabic{lemma}}
        \renewcommand{\thetheorem}{\Alph{appendixc}.\arabic{theorem}}
        \renewcommand{\thedefinition}{\Alph{appendixc}.\arabic{definition}}
        \renewcommand{\thecorollary}{\Alph{appendixc}.\arabic{corollary}}
        \renewcommand{\theequation}{\Alph{appendixc}.\arabic{equation}}
        \noindent{\tenbf Appendix \theappendixc #1}\par\vspace{5pt}}
\newcommand{\subappendix}[1] {\vspace{12pt}
        \refstepcounter{subappendixc}
        \noindent{\bf Appendix \thesubappendixc. {\kern1pt \bfit #1}}
	\par\vspace{5pt}}
\newcommand{\subsubappendix}[1] {\vspace{12pt}
        \refstepcounter{subsubappendixc}
        \noindent{\rm Appendix \thesubsubappendixc. {\kern1pt \tenit #1}}
	\par\vspace{5pt}}

\newcommand{\textlineskip}{\baselineskip=13pt}
\newcommand{\smalllineskip}{\baselineskip=10pt}

\def\eightcirc{
\begin{picture}(0,0)
\put(4.4,1.8){\circle{6.5}}
\end{picture}}
\def\eightcopyright{\eightcirc\kern2.7pt\hbox{\eightrm c}} 

\newcommand{\copyrightheading}[1]
	{\vspace*{-2.5cm}\smalllineskip{\flushleft
	{\footnotesize Modern Physics Letters A, #1}\\
	{\footnotesize $\eightcopyright$\, World Scientific Publishing
	 Company}\\
	 }}

\newcommand{\publisher}[2]{{\begin{center}\footnotesize\smalllineskip 
	Received #1\\
	Revised #2
	\end{center}
	}}

\def\abstracts#1#2#3{{
	\centering{\begin{minipage}{4.5in}\footnotesize\baselineskip=10pt
	\parindent=0pt #1\par 
	\parindent=15pt #2\par
	\parindent=15pt #3
	\end{minipage}}\par}} 

\def\keywords#1{{
	\centering{\begin{minipage}{4.5in}\footnotesize\baselineskip=10pt
	{\footnotesize\it Keywords}\/: #1
	 \end{minipage}}\par}}

\renewenvironment{thebibliography}[1]
	{\frenchspacing
	 \ninerm\baselineskip=11pt
	 \begin{list}{\arabic{enumi}.}
        {\usecounter{enumi}\setlength{\parsep}{0pt}     
	 \setlength{\leftmargin 12.7pt}{\rightmargin 0pt} 
         \setlength{\itemsep}{0pt} \settowidth
	{\labelwidth}{#1.}\sloppy}}{\end{list}}

\newcounter{itemlistc}
\newcounter{romanlistc}
\newcounter{alphlistc}
\newcounter{arabiclistc}

\newcommand{\fcaption}[1]{
        \refstepcounter{figure}
        \setbox\@tempboxa = \hbox{\footnotesize Fig.~\thefigure. #1}
        \ifdim \wd\@tempboxa > 5in
           {\begin{center}
        \parbox{5in}{\footnotesize\smalllineskip Fig.~\thefigure. #1}
            \end{center}}
        \else
             {\begin{center}
             {\footnotesize Fig.~\thefigure. #1}
              \end{center}}
        \fi}

\newcommand{\tcaption}[1]{
        \refstepcounter{table}
        \setbox\@tempboxa = \hbox{\footnotesize Table~\thetable. #1}
        \ifdim \wd\@tempboxa > 5in
           {\begin{center}
        \parbox{5in}{\footnotesize\smalllineskip Table~\thetable. #1}
            \end{center}}
        \else
             {\begin{center}
             {\footnotesize Table~\thetable. #1}
              \end{center}}
        \fi}

\def\@citex[#1]#2{\if@filesw\immediate\write\@auxout
	{\string\citation{#2}}\fi
\def\@citea{}\@cite{\@for\@citeb:=#2\do
	{\@citea\def\@citea{,}\@ifundefined
	{b@\@citeb}{{\bf ?}\@warning
	{Citation `\@citeb' on page \thepage \space undefined}}
	{\csname b@\@citeb\endcsname}}}{#1}}

\newif\if@cghi
\def\cite{\@cghitrue\@ifnextchar [{\@tempswatrue
	\@citex}{\@tempswafalse\@citex[]}}
\def\citelow{\@cghifalse\@ifnextchar [{\@tempswatrue
	\@citex}{\@tempswafalse\@citex[]}}
\def\@cite#1#2{{$\null^{#1}$\if@tempswa\typeout
	{IJCGA warning: optional citation argument 
	ignored: `#2'} \fi}}

\def\pmb#1{\setbox0=\hbox{#1}
	\kern-.025em\copy0\kern-\wd0
	\kern.05em\copy0\kern-\wd0
	\kern-.025em\raise.0433em\box0}

\def\fnt#1#2{\footnotetext{\kern-.3em
	{$^{\mbox{\scriptsize #1}}$}{#2}}}

\def\fpage#1{\begingroup
\voffset=.3in
\thispagestyle{empty}\begin{table}[b]\centerline{\footnotesize #1}
	\end{table}\endgroup}

\def\runninghead#1#2{\pagestyle{myheadings}
\markboth{{\protect\footnotesize\it{\quad #1}}\hfill}
{\hfill{\protect\footnotesize\it{#2\quad}}}}
\font\tenrm=cmr10
\font\tenit=cmti10 
\font\tenbf=cmbx10
\font\bfit=cmbxti10 at 10pt
\font\ninerm=cmr9

\font\eightrm=cmr8

\def\qed{\hbox{${\vcenter{\vbox{			
   \hrule height 0.4pt\hbox{\vrule width 0.4pt height 6pt
   \kern5pt\vrule width 0.4pt}\hrule height 0.4pt}}}$}}

\renewcommand{\thefootnote}{\fnsymbol{footnote}}	

\def\lesssim{\mathrel{\hbox{\rlap{\hbox{\lower4pt\hbox{$\sim$}}}\hbox{$<$}}}}
\def\gtrsim{\mathrel{\hbox{\rlap{\hbox{\lower4pt\hbox{$\sim$}}}\hbox{$>$}}}}

\def\pbar{{\overline{p}}}
\def\tpbar{\tau_{\overline{p}}}
\def\taup{\tau_p}
\def\Gampbar{\Gamma_{\pbar}}
\def\Gamp{\Gamma_p}
\def\overGam{\overline\Gamma}
\def\planck{M_{\rm Planck}}

\def\hrho{\hat\rho}
\def\hHam{\hat{\bf H}}
\def\hMass{\hat{\bf M}}
\def\hGamma{\hat{\bf\Gamma}}
\def\hOne{\hat{\bf 1}}
\def\hsigma{\hat{\bf\sigma}}
\def\vareps{\varepsilon}

\begin{document}
\setlength{\textheight}{7.7truein}  

\runninghead{CPT-- and B--Violation: The $p-\overline{p}$ Sector}{CPT-- 
and B--Violation: The $p-\overline{p}$ Sector}

\normalsize\textlineskip
\thispagestyle{empty}
\setcounter{page}{1}

\copyrightheading{}			

\vspace*{0.88truein}

\fpage{1}
\centerline{\bf $CPT$-- and $B$--VIOLATION: THE $p-\overline{p}$ SECTOR
\footnote[1]{\rm
Based on a contribution to the $CPT$ and Lorentz Symmetry Meeting (CPT98),
Indiana University, November 1998.}}
\baselineskip=13pt
\vspace*{0.37truein}
\centerline{\footnotesize DALLAS C. KENNEDY}
\baselineskip=12pt
\centerline{\footnotesize\it Department of Physics, University of Florida,}
\baselineskip=10pt
\centerline{\footnotesize\it Gainesville FL 32611-8440 USA}
\baselineskip=10pt
\centerline{\footnotesize\it E-mail: kennedy@phys.ufl.edu}
\vspace*{10pt}

\publisher{(accepted)}{(vol 14(13) (1999) 849)}

\vspace*{0.21truein}
\abstracts{The $CPT$ symmetry of relativistic quantum field theory 
requires the total lifetimes of particles and antiparticles be equal.  
Detection of $\pbar$ lifetime shorter than $\taup\gtrsim$ ${\cal O}(10^{32})$ 
yr would signal breakdown of $CPT$ invariance, in combination with 
$B$--violation.  The best current limit on $\tpbar$, inferred 
from cosmic ray measurements, is about 
one Myr, placing lower limits on $CPT$--violating scales that depend on the 
exact mechanism.  Paths to $CPT$ breakdown within and outside ordinary 
quantum mechanics are sketched.  Many of the limiting $CPT$--violating scales
in $\pbar$ decay lie within the weak--to--Planck range.}{}{}

\vspace*{10pt}

\textlineskip			
\vspace*{12pt}			

The $CPT$ symmetry of local relativistic quantum field theory (LRQFT)
has been tested in a number of elementary systems, in some cases to great
accuracy.~\cite{kkbar,anommagmom,browne}  
$CPT$ conservation rests on Lorentz invariance, locality, microcausality, and 
uniqueness of the vacuum.\cite{cptthm}
With no obvious evidence to the contrary, we assume $CPT$ symmetry
to be good.  But, as in the case of other discrete symmetries, the
search for small, non-vanishing exceptions is important in its own right
and as a check for residual forces outside the Standard Model.

$CPT$ symmetry requires that the properties of matter
and antimatter particles (charge, mass, magnetic moment) be related
by charge conjugation.
Less precise tests of $CPT$ have also been conducted with other 
systems:~\cite{cptother,cptppbar} here the comparison of the proton and 
antiproton decay lifetimes, $\taup$ and $\tpbar$, is presented.  The masses
and electromagnetic properties of $p$ and $\pbar$ are still assumed
identical for simplicity and charge $(Q)$ conservation unbroken.
This test of $CPT$ invariance is the first
to be directly combined with baryon number $(B)$ violation.

\setcounter{footnote}{0}
\renewcommand{\thefootnote}{\alph{footnote}}

\section{Theoretical considerations}
\noindent

Such properties as intrinsic decay lifetimes cannot be
limited with artificially produced antimatter at a level remotely approaching
that possible with matter, as there is never enough antimatter available
for a long enough time.
As a result, the decay lifetime of single antiprotons has been limited
by laboratory measurements to no more than $\tpbar >$ 3.4 
months~\cite{cptpbarlab}; while the decay of stored accelerator $\pbar$'s
is somewhat more restricted, $\tpbar/B(\pbar\rightarrow \mu^-\gamma ) >$  
$0.05\times 10^6$ yr (0.05 Myr) for the most sensitive exclusive 
mode.~\cite{cptAPEX}

With timescales much longer than possible in the laboratory, astrophysical 
antimatter processes provide some compensation
for antimatter measurements.
The resulting antimatter abundances can be inferred from cosmic ray
measurements.  The intrinsic timescale of cosmic antiprotons, produced
by secondary processes and stored temporarily in our Galaxy, is approximately
10 Myr.~\cite{Stephens,second_Webber}

Since $\taup\gtrsim$ ${\cal O}(10^{32})$ yr,~\cite{cptppbar} 
a detected $\tpbar\lesssim$ $10^7$ yr would be {\em prima facie} evidence of 
$CPT$ violation, as otherwise $\Gampbar$ = $\Gamp$ automatically.  
A relevant cosmic ray limit has been recently derived:~\cite{GeerKennedy} 
$\tpbar >$ 0.8 Myr\ (90\% C.L.).

\subsection{$CPT$ violation in standard quantum 
mechanics}\label{subsec:cptviol}

Without modifying basic quantum dynamics, $CPT$--violating effects can be 
introduced 
as small QFT modifications to the Standard Model (SM).~\cite{cptviolKostel}  
Such extensions can violate one or more of any of the four preconditions for
$CPT$ invariance, but the easiest to implement is Lorentz 
violation.~\cite{cptviolBigi}

If we assume that only one new, large scale, $M_X$, is involved in the $CPT$
violation, then the details of the interactions can be bypassed with
naive dimensional analysis.  On dimensional grounds, a QFT operator
in an extended SM Lagrangian density with canonical (mass) dimension $n >$
4 must be suppressed by $M^{4-n}_X$.  The resulting decay rate must 
be of the form
\begin{eqnarray}
\Gampbar\sim m_p\cdot [m_p/M_X]^{2n-8}\quad .
\end{eqnarray}
Ignoring dimensionless factors and neglecting $\Gamp$ compared to $\Gampbar$,
the new $CPT$--violating scale $M_X$ is then related to the $\pbar$ 
lifetime by $M_X = m_p\cdot [m_p\tpbar]^{1/(2n-8)}.$
Current limits on $\tpbar$ result in scales $M_X\lesssim$ $\planck$
(see section~\ref{sec:limits} below).

\subsection{$CPT$ violation via extensions of quantum 
mechanics}\label{subsec:cptviolNSQM}

An additional path to $CPT$ violation appears if we modify standard
quantum mechanics.  The most general description of a mixed quantum state is
given by the density matrix $\hrho (t)$ evolving via the Liouville-von Neumann
(LvN) equation:
\begin{eqnarray}
i{d\hrho (t)\over dt} = [\hHam, \hrho ]\quad ,
\label{eq:wigner}
\end{eqnarray}
where the Hamiltonian $\hHam$ is usually assumed Hermitian.  Unitarity  
requires ${\rm Tr}(\hrho )$ = 1, and equation~(\ref{eq:wigner}) preserves that
condition.  Irreversible decay can be incorporated by dropping the 
hermiticity of $\hHam$ and writing $\hHam$ = $\hMass - i\hGamma /2$, where
$\hMass$ and $\hGamma$ are Hermitian and positive semi-definite.  Unitarity
is no longer preserved by the LvN equation, and ${\rm Tr}(\hrho )$
decreases in time.  Alternatively, the {\em entropy} $S(t)$ increases:
$S = - {\rm Tr}(\hrho\ln\hrho ),$ ${\dot S}\ge 0.$

Inspired by his discovery of quantum black hole radiation, 
Hawking proposed to extend quantum mechanics~\cite{infolossHawking} 
with new terms in~(\ref{eq:wigner}) that
transform pure to mixed states.
Such a non-standard quantum mechanics (NSQM) was later shown by Page to
imply $CPT$ violation and by Banks, Susskind, and 
Peskin to create a conflict between locality and Lorentz invariance 
(energy--momentum conservation).~\cite{infolossOthers}  
The additional  terms in question add
to other possible $CPT$--violating effects that can already occur in
ordinary quantum field theory.
This NSQM formalism has been applied to neutral kaons~\cite{nsqmKKbar}
and solar neutrinos.~\cite{nsqmSNUs}

A similar approach can be taken with the $p-\pbar$ system, working in 
a two-dimensional space, $(p, \pbar)$, with a density matrix $\hrho$.
Decompose $\hrho$ in a general basis:
$\hrho = \rho^0\hOne + \rho^i\hsigma^i,$
and $\hMass$ and $\hGamma$ similarly, where $\hOne$ is the identity and
$\hsigma^i$ the Pauli matrices.  Take the most general linear extension
of the LvN equation in this space:
\begin{eqnarray}
{d\hrho\over dt} = 2\epsilon^{ijk}M^i\rho^j\hsigma^k -
\Gamma^0\hrho - \Gamma^i(\rho^0\hsigma^i+\rho^i\hOne ) -
h^{0j}\rho^j\hOne - h^{j0}\hsigma^j - h^{ij}\hsigma^i\rho^j~,
\label{eq:extendWigner}
\end{eqnarray}
with the summation convention assumed and the last three terms as the
NSQM extensions.  

The equation~(\ref{eq:extendWigner}) can be simplified with
reasonable assumptions:
absorbing as many terms as possible into shifts of $\hMass$ and $\hGamma ;$ 
equality of masses, $m_p$ = $m_\pbar ;$ and requiring ${\dot S}\ge$ 
0 (positivity).
These requirements eliminate $h^{0j}$ and $h^{j0}$ and imply 
$h^{ij}\ge$ 0.~\cite{nsqmKKbar}  We also assume
$\Delta Q = 0$, which further simplifies~(\ref{eq:extendWigner}) by
forcing all off-diagonal terms to zero and leaving only $h\equiv$ $h^{33}$
(not to be confused with Planck's constant).  The only other $CPT$--violating
effect arises from $\Delta\Gamma\equiv$ $\Gampbar - \Gamp$, the component
of $\hGamma$ proportional to $\hsigma^3$.  The density matrix $\hrho$ has
only diagonal components $\rho_+$ and $\rho_-$, and~(\ref{eq:extendWigner})
can be reduced to two coupled equations:
\begin{eqnarray}
{\dot\rho_\pm (t)} = -[\overGam\mp\Delta\Gamma /2 + h/2]\cdot\rho_\pm
+ h\rho_\mp/2\quad ,
\label{eq:extendReduce}
\end{eqnarray}
with $\overGam\equiv$ $(\Gampbar +\Gamp )/2$.  Note that baryon number $B$
is the operator $\hat{\bf B}$ = $\hsigma^3$, and its expectation value is
\begin{eqnarray}
\langle B(t)\rangle = {\rm Tr}[\hat{\bf B}\hrho (t)]\quad .
\end{eqnarray}

Subject~(\ref{eq:extendReduce}) to eigenmode analysis and 
take ${\rm Tr}[\rho (0)]$ = 1.  
The extensions of~(\ref{eq:extendWigner}) imply nothing about locality,
and these terms in general are nonlocal and/or acausal.  So assume an initial
state of matter consistent with typical laboratory conditions, but also
with the presumed state of the Universe: $\rho_+(0)$ = $1-\varepsilon$,
$\rho_-(0)$ = $\varepsilon$, with $\varepsilon\rightarrow$ 0.
The effective decay rates, if small, are then
\begin{eqnarray}
\Gamma^{\rm eff}_p     & = & -{\dot\rho}_+(0) = (1-\vareps )\Gamp + 
h/2\rightarrow\Gamp + h/2\quad ,\\ \nonumber
\Gamma^{\rm eff}_\pbar & = & -{\dot\rho}_-(0) = \vareps\Gampbar + 
(1+2\vareps )\cdot h/2\rightarrow h/2\quad .
\end{eqnarray}
Note that the NSQM evolution corrections are partly independent of the $\pbar$
concentration $\vareps$.  The {\em relative} $\pbar$ decay rate 
$-{\dot\rho}_-(0)/\vareps$ is enhanced by the small $\pbar$ concentration in
astrophysical situations:~\cite{Stephens,second_Webber,GeerKennedy}
$\vareps^{-1}\sim 10^{4-5}$.

Again assume only one new, large $CPT$--violating scale, $M_Y,$ that gives 
rise to $h$ in this $B$-violating sector; naive dimensional analysis again
indicates $h$ = $m^{k+1}_p/M^k_Y$, $k\ge$ 1, ignoring dimensionless factors.
Then $M_Y = m_p\cdot [m_p/h]^{1/k}.$
For higher--dimensional (larger $k)$ cases in the following 
section, the limiting scales again fall in the 
weak--to--Planck range.

\section{Implications for $CPT$ violation and associated 
scales}\label{sec:limits}
\noindent

Approximate limits on $M_X,$ $h,$ and $M_Y$ can be inferred from the lifetime
limits $\taup\gtrsim 10^{32}$ yr and $\tpbar\gtrsim 10^{7}$ 
yr.~\cite{cptppbar,GeerKennedy}  The
lower limits on $M_X$ are listed in the left two columns of Table~1.  
Note that the {\em largest}
$M_X$ is about $\planck$, while the next largest lies within the 
``intermediate'' range associated with left-right unification
and Peccei-Quinn symmetry breaking.~\cite{intermedUnifPQ}  
From higher-dimensional operators come lower scales close to the weak 
scale.  These may seem implausible, but might fit into ``large-radius''
$D\ge$ 4--brane compactifications with macro- or mesoscopic modifications 
to gravity and TeV--scale unification.~\cite{largeRbranes}

Now consider the measured $B$--violation limits on NSQM extensions.
Here it is $\taup ,$ not $\tpbar ,$ that restricts $h$, as $\taup$ is so 
much better constrained: $h\lesssim$
$10^{-55}$ eV!  But the two right columns of Table~1 show
that considering the
NSQM term $h$ in terms of naive dimensional analysis results, for
higher--dimensional operators, 
$k\ge$ 4, in scales $M_Y$ lower than $\planck$ and
hence not so implausible.  The $k$ = 4 case falls close to the standard
grand unification scale; the $k$ = 5 and 6 cases are in the ``intermediate''
range again.

\begin{table}[t]
\caption{$CPT$-- and $B$--violating scale limits from $p$ lifetime $\taup$ 
= $10^{32}$ yr and $\pbar$ lifetime $\tpbar$ = $10^7$ yr (see text).
\label{tab:NSQMlimits}}
\vspace{0.2cm}
\begin{center}
\footnotesize
\begin{tabular}{|c|c|c|c|}
\hline
$n$ & $M_X$ (GeV) & $k$ & $M_Y$ (GeV) \\
\hline
 5 & $2\times 10^{19}$ & 1 & $9\times 10^{63}$\\
 6 & $4\times 10^9$ &    2 & $9\times 10^{31}$\\ 
 7 & $3\times 10^6$ &    3 & $2\times 10^{21}$\\
 8 & $6\times 10^4$ &    4 & $9\times 10^{15}$\\
 9 & $7\times 10^3$ &    5 & $6\times 10^{12}$\\
10 & $2\times 10^3$ &    6 & $4\times 10^{10}$\\
\hline
\end{tabular}
\end{center}
\end{table}

\section{Summary and Outlook}
\noindent

$CPT$ violation, in the context of the $B$--violating $p$ and $\pbar$
decays, can be limited with the best current limits on $\taup$
(laboratory) and $\tpbar$ (cosmic rays), providing a test of $CPT$
invariance different from non-baryonic limits.  Assuming
one new scale $M_X$ or $M_Y$ associated with the $CPT$-- and $B$--violating
interactions, these limits imply lower bounds on $M_X$ or $M_Y$ that may, 
depending on the exact mechanism, fall in the usual weak--to--Planck 
range.

Improvements in laboratory and cosmic ray antiproton 
measurements~\cite{pbarFuture} 
will further probe the combination of $CPT$
and baryon number violation.
Unified string or brane theories, while preserving $CPT$
at the dynamical level, could break it spontaneously upon 
compactification,~\cite{cptstrings} while the full effect of quantum gravity
on fundamental symmetries remains an enigma.
Perhaps a symmetry such as $CPT$ today so 
obvious may fall, like parity and $CP,$ under the ax of an unexpected
experimental result.

\nonumsection{Acknowledgments}
\noindent
The author thanks Alan Kosteleck\' y of Indiana University for the opportunity 
to present this work at the CPT '98 Conference, and Roberto Floreanini of INFN 
(Trieste) for helpful comments on quantum evolution.  This letter is based
on work done in collaboration with Stephen Geer of Fermilab and
supported at Fermilab under grants U.S. DOE DE-AC02-76CH03000 and NASA 
NAG5-2788, and at the Univ. Florida, Institute for Fundamental
Theory, under grant U.S. DOE DE-FG05-86-ER40272.

\nonumsection{References}

\end{document}